\documentclass[reprint,nofootinbib,amsmath,amssymb,aps, pra]{revtex4-2}

\usepackage{graphicx}
\usepackage{dcolumn}
\usepackage{bm}
\usepackage{hyperref}
\usepackage{svg}
\usepackage{makecell}
\usepackage{empheq}
\usepackage{dsfont}
\usepackage{bbold}

\newcommand{\ve}[1]{{\mathbf #1}}

\newcommand{\be}{\begin{equation}}
\newcommand{\ee}{\end{equation}}

\begin{document}

\title{
Open and trapping channels in complex resonant media
}
	
\author{Romain Rescanieres}
\author{Romain Pierrat}
\author{Arthur Goetschy}
\affiliation{ESPCI Paris, PSL University, CNRS, Institut Langevin, 1 rue Jussieu, F-75005 Paris, France}

\begin{abstract}

We present a statistical study of the transmission and dwell-time matrices in disordered media composed of resonators, focusing on how frequency detuning influences their eigenvalue distributions. Our analysis reveals that the distribution of transmission eigenvalues undergoes a transition from a monomodal to a bimodal profile, and back to monomodal, as the frequency approaches the resonant frequency of the particles. Moreover, the distribution of dwell-time eigenvalues broadens significantly near resonance, with the longest lifetimes exceeding the median by several orders of magnitude. These results are explained by examining how frequency $\omega$ affects the transport mean free path of light, $\ell(\omega)$, and the energy transport velocity, $v_E(\omega)$, which in turn shape the observed distributions. We demonstrate the strong potential of wavefront shaping to enhance both transmission and energy storage in resonant disordered media. In the diffusive regime, where the system thickness $L$ exceeds the mean free path, both transmission and dwell time can be enhanced by a factor $\varpropto L/\ell(\omega) \gg 1$ when using wavefronts associated with the largest eigenvalues instead of plane waves. In the localized regime, the enhancements become $\varpropto Ne^{2L/\xi}$ for transmission and $\varpropto N\xi /L$ for dwell time, where $\xi$ is the localization length and $N$ is the number of controlled scattering channels. Finally, we show that employing high-$Q$ resonators instead of low-$Q$ ones increases energy storage within the medium by a factor of $\varpropto Q/k\ell(\omega)$, in both the diffusive and localized regimes.

\end{abstract}

\maketitle
	
\section{Introduction}
\label{sec:introduction}

In recent years, the development of wavefront shaping techniques has enabled the measurement of the scattering matrix in various complex media, such as disordered waveguides~\cite{shi_12, sarma_16, bender_22}, scattering slabs~\cite{popoff_2010, hsu_17}, multimode fibers~\cite{choi_12, ploschner_15, matthes_21}, and biological tissues~\cite{badon_20, lee_22}, which can support a large number of propagation channels. These advancements have shown that specific wavefronts can be focused deep within turbid systems or fully transmitted through materials that are otherwise opaque to plane waves~\cite{mosk_12, cao_22}. Additionally, by calculating the derivative of the scattering matrix with respect to frequency, it is possible to construct the Wigner-Smith operator and the associated dwell-time operator, whose eigenvalues represent the duration a quasi-monochromatic wave packet can spend within an open system~\cite{texier_16, rotter_17}. Using a spatial light modulator to generate the eigenstates of the dwell-time operator, it is thus in principle possible to optimize the energy stored in any arbitrary complex system~\cite{durand_19}.

So far, these properties have been established in non-resonant scattering materials. However, resonant media are now ubiquitous in nanophotonics and atomic optics. Technological advancements have enabled the creation of optically resonant systems structured either at the wavelength scale (such as photonic crystals and arrays of microcavities) or at a deep subwavelength scale (such as plasmonic nanoresonators and atomic arrays). Light-matter interactions in these systems are typically described in terms of collective modes rather than by their scattering matrices~\cite{yariv_99, goetschy_11, hsu_16, limonov_17, shahmoon_17, lalanne_18, chang_18}. Furthermore, the effect of coherent control via wavefront modulation on such media has been scarcely explored. Previous research has examined a few resonant elements embedded in otherwise non-resonant materials, such as using resonators to focus light deep into complex media without a guide star~\cite{durand_19, delHougne_21} or employing fluorescent probes to resolve the 3D profile of open channels in a non-resonant disordered slab~\cite{hong_18}. In contrast, this work addresses systems where all scattering units are resonant, aiming to understand the influence of local resonances on the properties of transmission and dwell-time matrices.

The impact of resonant scattering on simple plane wave propagation is relatively well understood, as demonstrated in a variety of disordered systems, ranging from cold atomic gases~\cite{labeyrie_03} to random dielectric materials supporting Mie resonances~\cite{sapienza_07}, as well as acoustical systems made of droplets~\cite{tallon_17} or Helmholtz resonators~\cite{lemoult_13}. Internal resonances are known to affect both stationary and dynamic transport properties. The ability to tune the refractive index near resonance allows for significant variations in the phase and group velocities that characterize the propagation of the ballistic wave~\cite{tallon_17, hau_99}. However, in a disordered scattering medium, most of the energy is carried by the diffusive component rather than the ballistic wave component. This diffusive component propagates at a speed $v_E \le c$ for all frequencies and can be dramatically slowed down near resonance~\cite{lagendijk_96}. Additionally, the scattering cross-section of the resonators increases significantly near resonance, reducing the transport mean free path $\ell$ and mean transmission coefficient $\left< T \right>$, while substantially increasing the time needed to cross the sample, termed the Thouless time $\tau_{\mathrm{Th}}$. The objective of this work is to extend the analysis beyond plane wave behavior to arbitrary wavefronts, whose propagation cannot be adequately explained by the traditional diffusion model, even when the sample thickness $L$ greatly exceeds $\ell$~\cite{akkermans_montambaux_07}.

To address this problem, we consider light wave propagation in an assembly of point-like resonators randomly distributed within a finite volume much larger than the wavelength. For convenience, this volume is placed in a waveguide, which limits the size of the studied matrices to the number of propagating modes in the waveguide. Sec.~\ref{trans_time_mat} focuses on establishing the coupled equations describing light-matter interactions and the procedure for constructing the transmission and reflection matrices from these equations. It also presents the results of numerical simulations of the transmission matrix, evaluated at different frequencies and across numerous disorder configurations. The statistical properties of transmission eigenvalues and eigenstates as a function of frequency $\omega$ are shown to depend crucially on the mean free path $\ell (\omega)$, allowing for the exploration of different propagation regimes (quasi-ballistic, diffusive, localized). In a symmetric fashion, Sec.~\ref{dwell_time_matrix} presents results related to the dwell-time operator, demonstrating that the behavior of eigenvalue distribution and eigenstate propagation can be understood through the interplay between the frequency-dependent mean free path $\ell(\omega)$ and energy velocity $v_E(\omega)$. Sec.~\ref{conclusion} provides a summary of the results, emphasizing the significant advantages of wavefront shaping in achieving much higher transmission and energy storage compared to simple plane wave propagation, both in the diffusive and localized regimes.

\section{Transmission matrix and open channels}
\label{trans_time_mat}

\subsection{Transmission and reflection matrices for resonant scattering media}

\begin{figure}	
\centering
\includegraphics[width = 0.9 \linewidth]{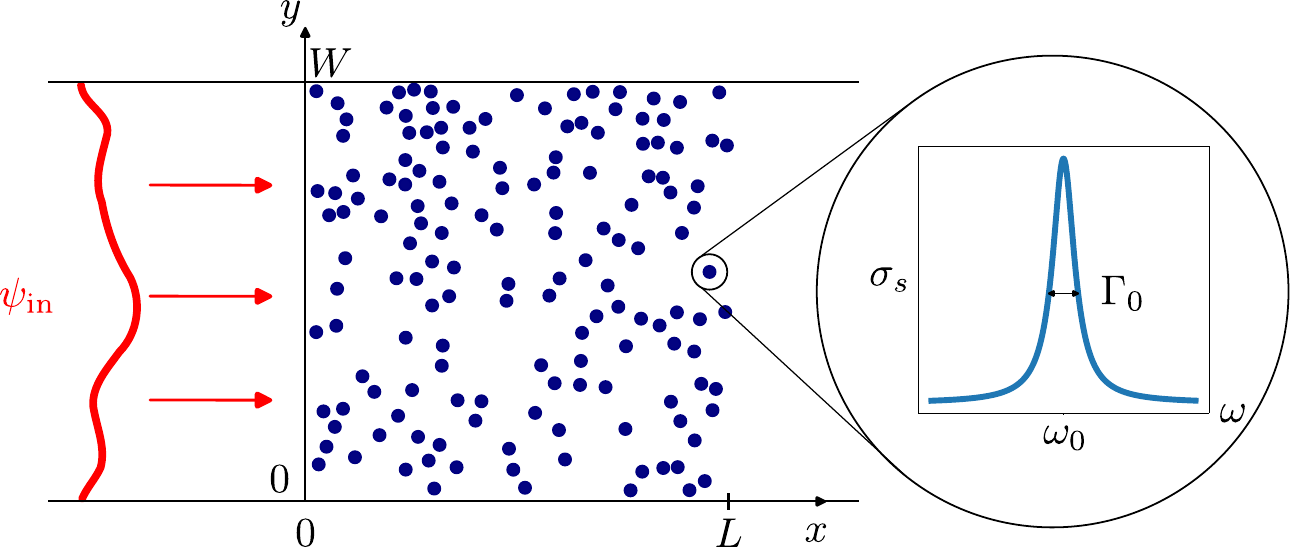}
\caption{Schematic representation of the system under study. $N_s$ resonant scatterers are uniformly distributed in a region of size $L\times W$ of a  2D waveguide. An incident field $\psi_{\mathrm{in}}$ at frequency $\omega$ is sent onto the scattering region from the left. The resonant scatterers, here represented with a finite size, will be taken to be point-like in the rest of this paper. Their quality factor is $Q = \omega_0/(\Gamma_0/2)$. The inset shows the scattering cross-section $\sigma_s = k^3 \vert \alpha(\omega) \vert^2/4$ of a single resonator in free space, with polarizabilty $\alpha(\omega)$ given by Eq.~\eqref{EqAlphaFreeSpace}.  }
\label{fig:system}
\end{figure}

The system under study consists in an assembly of $N_s$ identical point-like resonant scatterers of polarizability $\alpha(\omega)$, randomly distributed in a two-dimensional waveguide of thickness $L$ and width $W$, as sketched in Fig. \ref{fig:system}. It is illuminated from the left by waves of frequency $\omega = ck$, polarized along the $z$-direction. In order to compute the transmission matrix and dwell-time operator, we need to evaluate the scalar Green's function of the wave equation, $G(\ve{r}, \ve{r'}, \omega)$, for a source located at any point $\ve{r}'=(x', y')$ on the front surface of the medium and an observation point $\ve{r} =(x,y)$ located either on the back or on the front surface.
In Appendix~\ref{app:couple_dipole}, we demonstrate that, in the presence of the waveguide, the coupled dipole equations governing the evolution of \( G(\ve{r}, \ve{r'}, \omega) \) can be expressed in the following form:
\begin{gather}
G(\ve{r}, \ve{r'}, \omega)\!=\!\tilde{G}_0(\ve{r}, \ve{r'}, \omega) 
\!-k^2 \!  \sum_{i=1}^{N_s}  \tilde{G}_0(\ve{r},\ve{r}_i,\omega)\tilde{\alpha} (\ve{r}_i, \omega)\tilde{\psi}_i,
\label{EqCoupleDipole3}
\\
\tilde{\psi}_i = \tilde{G}_0(\ve{r}_i,\ve{r}',\omega)
 -k^2 \sum_{\substack{j=1\\j\neq i}}^{N_s} \tilde{G}_0(\ve{r}_i,\ve{r}_j,\omega) \tilde{\alpha} (\ve{r}_j, \omega) \tilde{\psi}_j.
\label{EqCoupleDipole4}
\end{gather}
Here, $\tilde{G}_0(\ve{r}, \ve{r'}, \omega)$ is the retarded Green's function of the wave equation in the empty waveguide, which solves $(\nabla^2 +k^2 +i0^+)\tilde{G}_0(\ve{r}, \ve{r'}, \omega)=\delta(\ve{r}- \ve{r'})$. The presence of the waveguide imposes Dirichlet boundary conditions in the $y$-direction, yielding
\be
\tilde{G}_0(\ve{r},\ve{r}',\omega)  =\frac{1}{W} \sum_{n>0} \sin\left(\frac{n\pi y}{W}\right)\sin\left(\frac{n\pi y'}{W}\right)  \frac{e^{ik_{n}|x-x'|}}{ik_{n}},
\label{EqGreenGuide}
\ee 
where $k_{n}=k\sqrt{1-(n\pi /kW)^2}$. In Eqs.~\eqref{EqCoupleDipole3} and \eqref{EqCoupleDipole4}, $\tilde{\psi}_i$ is the effective field exciting the scatterer located at the position $\ve{r}_i$. Note that the expression of the polarizability $\tilde{\alpha} (\ve{r}_i, \omega)$ of this scatterer is modified compared to its free space value $\alpha(\omega)$. It is given by (see Appendix~\ref{app:couple_dipole})
\be
\frac{1}{\tilde{\alpha} (\ve{r}_i, \omega)} = \frac{1}{\alpha(\omega)} + k^2\Delta G_0(\ve{r}_i, \omega),
\label{EqRenormalizedAlpha}
\ee
where $\Delta G_0(\ve{r}_i, \omega)=\tilde{G}_0(\ve{r}_i,\ve{r}_i,\omega) -G_0(\ve{r}_i,\ve{r}_i,\omega)$, with $G_0$ the retarded free space Green's function.
For $z$-polarized waves in 2D, the bare polarizability reads
\be
\alpha(\omega)= \frac{-4c^2\Gamma_0/\omega_0}{\omega^2-\omega_{0}^2 + i\Gamma_0\omega^2/\omega_0},
\label{EqAlphaFreeSpace}
\ee
where $\omega_0$ and $\Gamma_0$ are respectively the resonant frequency and linewidth in free space (see inset of Fig. \ref{fig:system}). This expression satisfies the condition 
$\mathrm{Im}[1/\alpha (\omega)]=k^2\mathrm{Im}[G_0(\ve{r}_i,\ve{r}_i,\omega)]$, imposed by energy conservation of the scattering process in free space.
In addition,  $G_0(\ve{r}, \ve{r}',\omega)=-(i/4) H_0^{(1)}(k|\ve{r}-\ve{r}'|)$ , with $H_0^{(1)}$ the Hankel's function of the first kind of order $0$. The numerical procedure used to evaluate $\Delta G_0(\ve{r}_i, \omega)$ efficiently is detailed in Appendix~\ref{app:delta_G0}.

Consequently, replacing free space with a waveguide means not only replacing the free space retarded Green's function $G_0$ with $\tilde{G}_0$ in the coupled dipole equations, but also renormalizing the polarizability. This renormalization is essential to satisfy the energy conservation condition $\mathrm{Im}[1/\tilde{\alpha} (\ve{r}_i, \omega)]=k^2\mathrm{Im}[\tilde{G}_0(\ve{r}_i,\ve{r}_i,\omega)]$ in the waveguide. The local polarizability can also be expressed as
\be
\tilde{\alpha}(\ve{r}_i,\omega)= \frac{-4c^2\Gamma_0/\omega_0}{\omega^2-\tilde{\omega}_{0}(\ve{r}_i)^2 + i\tilde{\Gamma}_0(\ve{r}_i)\omega^2/\omega_0},
\ee
where $\tilde{\omega}_{0}(\ve{r}_i)^2 =\omega_0^2+4\omega^2\Gamma_0\mathrm{Re}\left[\Delta G_0(\ve{r}_i, \omega) \right]/\omega_0$ and $\tilde{\Gamma}_0(\ve{r}_i)=-4\Gamma_0\mathrm{Im}\left[\tilde{G}_0(\ve{r}_i,\omega) \right]$. In the waveguide, translation invariance is broken, so that both the Lamb-shift and decay rate, proportional to the local density of state, depend on the position $\ve{r}_i$ of the resonator.

We can express now the transmission matrix $\ve{t}(\omega)$  and reflection matrix $\ve{r}(\omega)$ of the disordered medium using a basis composed of the $N = \lfloor kW/\pi \rfloor $ propagating channels of the empty waveguide. These channels are represented by the functions $e^{ik_nx} \xi_n(y)$, with $\xi_n(y) = \sqrt{2/W}\sin(n\pi y/W)$. After normalizing each state by its flux, $\varepsilon_0ck_n/2 k$, the elements of the $N\times N$ transmission and reflection matrices can then be determined through the Fisher-Lee relations~\cite{fisher_lee_81}, 
\begin{align}
t_{mn}(\omega) &= 2i\sqrt{k_{m}k_{n}} \langle L,\xi_m | \mathbf{G}(\omega) |0,\xi_n\rangle, 
\label{EqTM}
\\
r_{mn}(\omega) &= -\delta_{mn} + 2i\sqrt{k_{m}k_{n}}\langle 0,\xi_m |\mathbf{G}(\omega) |0,\xi_n\rangle,
\label{EqRM}
\end{align}
where the Green's operator $\mathbf{G}(\omega)$ is defined through its real-space representation, $\langle \ve{r}| \mathbf{G}(\omega) |\ve{r}'\rangle = G(\ve{r},\ve{r}', \omega)$. The choice of normalization guarantees that the matrices $\ve{t}(\omega)$ and $\ve{r}(\omega)$ satisfy the condition 
\be
\ve{t}(\omega)^\dagger\ve{t}(\omega) + \ve{r}(\omega)^\dagger\ve{r}(\omega)= \mathbf{1} 
\ee
in the absence of absorption. This flux conservation condition is only satisfied  if the dressed polarizabilty $\tilde{\alpha}(\ve{r}_i, \omega)$ is used in the evaluation of the Green's operator; it is not satisfied if the bare polarizability $\alpha(\omega)$ is used instead. 

According to Eqs.~\eqref{EqCoupleDipole3} and~\eqref{EqCoupleDipole4}, we can write the Green's operator in the form 
\be
\mathbf{G}(\omega) = \mathbf{\tilde{G}_0}(\omega) + \mathbf{\tilde{G}_0}(\omega) \mathds{T}(\omega)\mathbf{\tilde{G}_0}(\omega),
\label{EqGreenT}
\ee
where the collective $\mathds{T}$ operator is defined as
\be
 \mathds{T}(\omega)=-k^2\sum_{i=1}^{N_s}\sum_{j=1}^{N_s}\left[\frac{\mathds{a}(\omega)}{\mathds{1}+k^2 \mathds{G}_0(\omega)\mathds{a}(\omega)}\right]_{ij}|\ve{r}_i \rangle \langle \ve{r}_j |.
\ee
Here, $\mathds{a}(\omega)$ and $\mathds{G}_0(\omega)$ are $N_s\times N_s$ matrices, defined as $\mathds{a}_{ij}(\omega)=\tilde{\alpha}(\ve{r}_i, \omega)\delta_{ij}$ and $\mathds{G}_{0,ij}(\omega)=(1-\delta_{ij})\tilde{G}_0(\ve{r}_i,\ve{r}_j,\omega).
$
Fisher-Lee relations and Eq.~\eqref{EqGreenT} reveal that the matrices  $\ve{t}(\omega)$ and $\ve{r}(\omega)$ can be  computed directly in the constant-flux basis from the combination of the $ \mathds{T}$ operator with the propagators $\langle x,\xi_m | \mathbf{\tilde{G}_0}(\omega) |0,\xi_n\rangle =  \delta_{mn} e^{ik_{m}x}/2ik_{m}$ and $\langle x,\xi_m | \mathbf{\tilde{G}_0}(\omega) |\ve{r}_i \rangle =  \xi_m(y_i)e^{ik_{m}\vert x - x_i\vert}/2ik_{m}$. The simulation results presented in the following sections were obtained using this procedure, whose simulation cost depends only on $N_s$ and $N$. It is in particular completely independent of the system length $L$. 

\subsection{Transmission eigenvalue distribution}

\begin{figure}
\centering
\includegraphics[width = 0.9\linewidth]{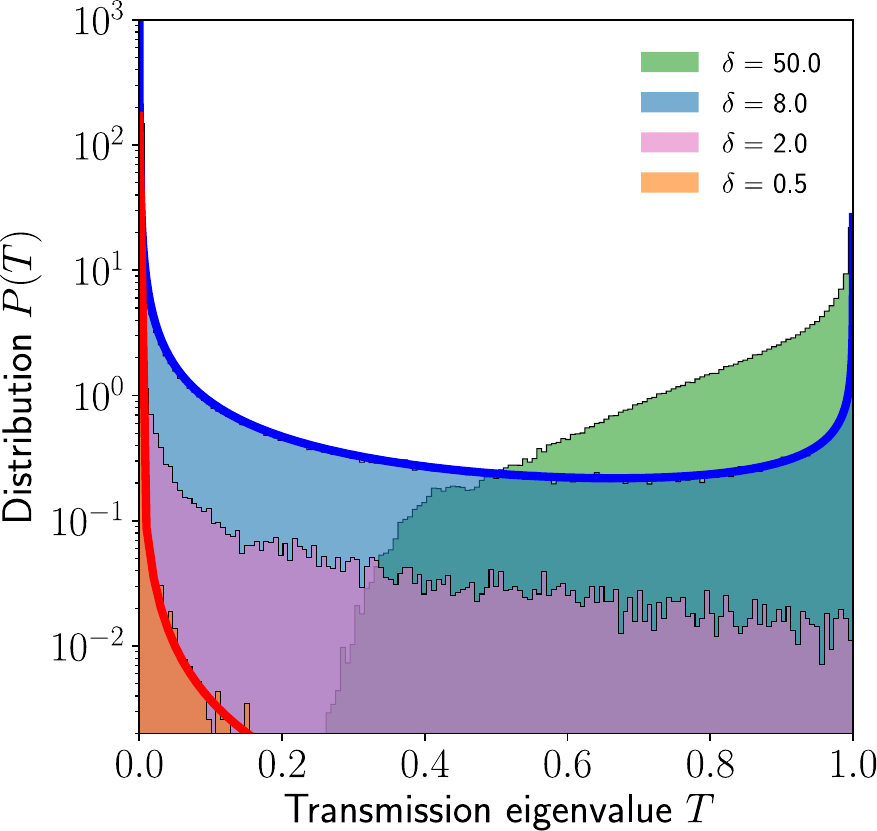}
\caption{Probability distribution of the eigenvalues $T_n$ of the matrix $\ve{t}^{\dagger}(\omega)\ve{t}(\omega)$. Different histograms correspond to different detuning values $\delta = 2(\omega-\omega_0)/\Gamma_0$, and therefore to different mean free paths (see Table \ref{tab:ex} for reference). Blue solid line is the analytical prediction in the diffusive regime given by Eqs.~\eqref{EqBimodal},~\eqref{EqMeanT} and~\eqref{EqMeanFreePath}. Red solid line corresponds to the analytical prediction in the localized regime, given by Eqs.~\eqref{EqPDFLoc} and~\eqref{EqPMax}, where $\xi = N\pi \ell(\omega)/2$ with $\ell(\omega)$ given by Eq.~\eqref{EqMeanFreePath}. All distributions are computed for 5760 disorder configurations with  $N_s = 2\times 10^4$ scatterers, with a quality factor $Q = \omega_0/(\Gamma_0/2) = 10^3$ in a waveguide of longitudinal size $k_0L = 5000$ and of transverse size $k_0W = 161$. The number of modes (which increases with the detuning) is $N= 53$ in the ballistic regime and $N= 51$ in the other regimes. }
\label{fig:eigen_transm}
\end{figure}

Our goal is to study the impact of the detuning $\delta=2(\omega-\omega_0)/\Gamma_0$ on the distribution of the eigenvalues $T_n$ of the matrix $\ve{t}(\omega)^\dagger \ve{t}(\omega)$. This distribution is defined as
\be
P(T)=\frac{1}{N} \left< \sum_{n=1}^N \delta(T-T_n)\right>.
\ee
In order to explore different propagation regimes by simply tuning the frequency, we focus our analysis to systems of width $W \lesssim N \ell(\omega)$, for which the localization length is $\xi \simeq \pi N \ell(\omega)/2$~\cite{beenakker_97}. Since $N\simeq kW/\pi$, this imposes that $k\ell(\omega) \gtrsim1$. In this regime, an estimate of the mean free path in a disordered medium of density $\rho=N_s/LW$ is given by~\cite{carminati_21,monsarrat_22}
\be
k\ell(\omega)=\frac{4}{\rho k^2 \vert \alpha(\omega)\vert^2} \underset{\delta \ll Q}{\simeq} \frac{(kW)(kL)}{4N_s}(\delta^2+1).
\label{EqMeanFreePath}
\ee 
Reaching the diffusive regime $\xi \gg L \gg \ell(\omega)$ requires a large number $N$ of propagating channels, while reaching the localized phase close to resonance ($ W \lesssim N \ell(\omega_0)  \lesssim L$) requires a number of scatters  $N_s$ such that $(kW)(kL) \gtrsim N_s\gtrsim (kW)^2$. All those requirements will be met in the following results. Figure~\ref{fig:eigen_transm} shows the distribution $P(T)$ at different detuning $\delta$, in a waveguide supporting $N\simeq 50$ channels ($k_0W=161$) filled with $N_s=2\times10^4$ scatterers of quality factor $Q=\omega_0/(\Gamma_0/2) =10^3$.  Each value of detuning corresponds to a different transport regime, as reported in Table~\ref{tab:ex}. 

\begin{table}[b!]
\begin{ruledtabular}
\begin{tabular}{ccccc}
\makecell{Transport\\regime} & Localized & \makecell{Quasi-\\localized} & Diffusive & \makecell{Quasi-\\ballistic} 
\\
\hline
\\
$\delta$ & $0.5$ & $2$ & $8$ & $50$ 
\\
\\
\hline
\\
$\frac{\ell}{L} \varpropto \frac{N}{N_s} (\delta^2+1)$ & $2.5 \cdot 10^{-3}$ & $1.0 \cdot 10^{-2}$ &  $1.3 \cdot 10^{-1}$ & $5.0$
 \\
 \\
\hline
\\
$\frac{\xi}{L} \varpropto \frac{N^2 }{N_s}(\delta^2+1)$ & $2.0 \cdot 10^{-1}$ & $8.0 \cdot 10^{-1}$ &  $1.0 \cdot 10^{1}$ & $4.2\cdot 10^{2}$
 \\
 \\
\end{tabular}
\end{ruledtabular}
\caption{Correspondence between detuning value $\delta$, mean free path $\ell$ and localization length $\xi$, normalized by the system thickness $L$. The values are given for the parameters used in the numerical simulations shown in Figs.~\ref{fig:eigen_transm} and~\ref{fig:eigen_DT}, namely $k_0L = 5000$, $k_0W = 161$ and $N_s = 20000$. For each detuning, the values of $\ell/L$ and $\xi/L$ determine the transport regime in the system.}
\label{tab:ex}
\end{table}

For large detuning ($\delta = 50$), the coupling between light and matter is weak. As a result, the mean free path, given by Eq.~\eqref{EqMeanFreePath}, exceeds the medium's thickness, causing most waves to undergo quasi-ballistic transport. Consequently, the transmission eigenvalue distribution $P(T)$, shown in Fig.~\ref{fig:eigen_transm} (shaded green), is peaked around 1, with a minimum eigenvalue $T_{\mathrm{min}}$ that remains strictly positive. This indicates that no wavefront is fully reflected by the scattering medium.
The precise value of $T_{\mathrm{min}}$ depends on the transverse width $W$, with $T_{\mathrm{min}} \to 0$ as $W \to \infty$. This is because the number of propagating channels, which travel along distances  $ L'_n = L/\sqrt{1-(n\pi /kW)^2} $, increases as $W$ grows, leading to more channels for which $ L'_n > \ell(\omega)$. We note that there is no theoretical prediction available to accurately describe $P(T)$ for $L \lesssim \ell(\omega)$. In particular, the prediction from the Dorokhov-Mello-Peyrera-Kumar (DMPK) equation is inaccurate (result not shown) because it relies on the assumption of scattering channel isotropy~\cite{dorokhov_84, mello_1988, beenakker_97}, which does not hold in the quasi-ballistic regime.

As $\omega$ progressively approaches $\omega_0$ ($\delta = 8$), light and matter increasingly couple, and the transport mean free path decreases. As a consequence, transmission eigenchannels cover the full range $T\in [0,1]$, with a distribution $P(T)$ that develops a second peak at $T=0$ (shaded blue in Fig.~\ref{fig:eigen_transm}). In the diffusive regime $\ell(\omega)\ll L \ll \xi$, the average transmission eigenvalue, $\left< T(\omega)\right> =\left< \mathrm{Tr} \left[\ve{t}(\omega)^\dagger \ve{t}(\omega)\right]\right>/N$, follows from the result of the diffusion equation for the mean intensity of light~\cite{akkermans_montambaux_07},
\be
\left< T(\omega)\right>=\frac{\pi\ell(\omega)/2}{L + \pi\ell(\omega)/2},
\label{EqMeanT}
\ee 
and the distribution $P(T)$ is very well described by the bimodal prediction~\cite{dorokhov_84, beenakker_97}
\be
P(T)= \frac{\langle T (\omega) \rangle}{2T\sqrt{1-T}},
\label{EqBimodal}
\ee
which supports a finite number $g= N \left< T(\omega)\right> \gg 1$ of open channels with transmission close to unity~\cite{imry_86}, whereas the medium is opaque for random wavefronts and plane waves ($\langle T (\omega) \rangle \ll 1 $).  The blue solid line  in Fig.~\ref{fig:eigen_transm} corresponds to the prediction given by Eqs.~\eqref{EqMeanFreePath}, ~\eqref{EqMeanT}, and~\eqref{EqBimodal}. Since the result in Eq.~\eqref{EqBimodal} can be derived using the DMPK approach  -- a scaling analysis of transmission that presumes a characteristic length scale, $\ell(\omega)$,  linked to a linearly inverse decrease in conductance with system size -- it is unsurprising that this result holds for the resonant media considered here, provided that $\ell(\omega)$ is appropriately calculated. 

\begin{figure*}[t!]
\centering
\includegraphics[width = 0.85\linewidth]{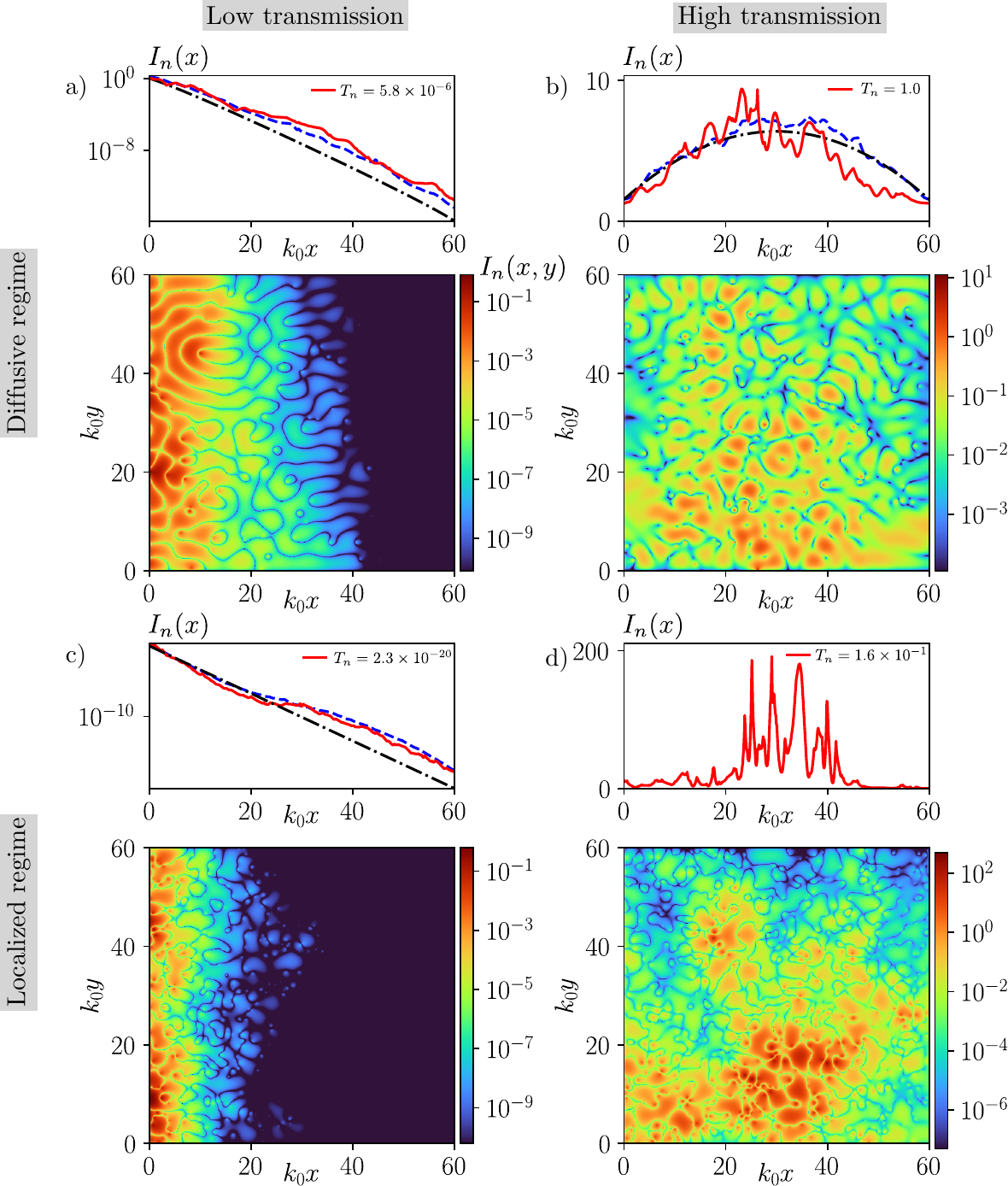}
\caption{Intensity maps $I_n(x,y)=|\psi_n(x, y)|^2$ and integrated intensity profiles $I_n(x)=\int_0^W \mathrm{d}y I_n(x,y)$ for typical incident wavefronts, chosen as different eigenstates $V_n$ of the matrix $\mathbf{t}^{\dagger}(\omega)\mathbf{t}(\omega)$, in both the diffusive (first row) and localized regimes (second row). The left (right) column corresponds to the propagation of a low (high) transmission eigenstate in each regime. For clarity, the results correspond to square systems with $k_0L = k_0W = 60$, at $\delta = 0$ and $Q=10^3$. In each regime, the number of scatterers is adjusted so that the distribution of transmission eigenvalues qualitatively resembles Fig.~\ref{fig:eigen_transm}: $N_s = 120$ in the diffusive regime and $N_s = 960$ in the localized regime. Solid red lines in $I_n(x)$ represent the specific disorder realization associated with the intensity map below, while blue dashed lines show the average intensity over 30 disorder configurations. Black dash-dotted lines in the diffusive regime correspond to Eq.~\eqref{EqOpen} and~\eqref{EqMeanFreePath} for the open channel and to Eq.~\eqref{EqClosed} for the closed channels. In the localized regime, numerical results for closed channel profile $\left<I_n(x)\right>$  are compared with the exponential profile $e^{-\ln(4/T_n)x/L}$.
}
\label{fig:maps_T}
\end{figure*}

When the localization length equals the system length ($\delta = 2$), the number of open channels reduces to $g \simeq 1$, causing the distribution $P(T)$ to lose its peak at $T = 1$ (shown in pink in Fig.~\ref{fig:eigen_transm}). At $\delta = 0.5$, where $\xi \ll L$, most transmission eigenvalues approach zero (see orange histogram in Fig.~\ref{fig:eigen_transm}). More specifically, for each realization of disorder, all eigenvalues $T_n$ are much smaller than the largest one. Consequently, the eigenvalue distribution takes the form
\be
P(T)\simeq \frac{N-1}{N}\delta(T) + \frac{1}{N}P_{\mathrm{max}}(T), 
\label{EqPDFLoc}
\ee
where $P_{\mathrm{max}}(T)$ is the distribution of $T_\mathrm{max}=\mathrm{max}(T_n)$. The largest eigenvalue, which exhibits significant fluctuations, follows an approximately log-normal distribution with $\mathrm{Var}[\ln(T_\mathrm{max})]=-2\left< \ln(T_\mathrm{max})\right>=4L/\xi \gg 1$. Its distribution can be closely approximated by
\be
P_{\mathrm{max}}(T)= C(\xi)\sqrt{\ln\left(\frac 4 T \right)}\frac{e^{-\frac{\xi}{2L}\left[\frac{\ln(T/4)}{2}+\frac{L}{\xi}\right]^2}}{T},
\label{EqPMax}
\ee
where  $C(\xi)$ is a normalization factor that depends on $\xi$. This expression is obtained from Refs.~\cite{gopar_10, abrikosov_81} in the limit of vanishing conductance $g$. The distribution $P(T)$ given by Eqs.~\eqref{EqPDFLoc} and~\eqref{EqPMax} is in very good agreement with our numerical results (red solid line in  Fig.~\ref{fig:eigen_transm}). It implies in particular that
\be
\left<T \right> \simeq \kappa \frac{e^{-L/2\xi}}{N} \ll \left<T_\mathrm{max} \right> \simeq\kappa e^{-L/2\xi} \ll \mathrm{max}(T_\mathrm{max}) \simeq 1,
\nonumber
\ee
where $\kappa = C(\xi) [2\sqrt{\pi}\mathrm{erfc}(\ln(4)^{1/2})+\ln4]$, with $\mathrm{erfc}$ the complementary error function. 
For a finite number $N_r$ of disorder realizations, a rough estimate of the largest accessible eigenvalue, $T_\mathrm{Max}=\mathrm{max}(T_\mathrm{max})$, is given by the solution of $NN_r\int_{T_\mathrm{Max}}^1\mathrm{d}T p(T)=1$.  In the simulations at $\delta=0.2$ presented in Fig.~\ref{fig:eigen_transm}, where $L/\xi=5$, $N=51$, and $N_r=5760$, we find $\langle T \rangle \simeq 1.3\times 10^{-4}$, $\langle T_{\mathrm{max}}\rangle \simeq 7 \times 10^{-3}$ and $T_\mathrm{Max} \simeq 0.2$. This shows that the use of wavefront shaping can help to gain orders of magnitude in transmission passing from an incident plane-wave to the maximally transmitted eigenchannel.

\subsection{Eigenstate propagation}

The eigenstates of $\ve{t}(\omega)^\dagger \ve{t}(\omega)$ are of the form $V_n(x,y)=\sum_{p=1}^N c_p^{(n)} \sqrt{k/k_p} e^{ik_{p}x} \xi_p(y)$, with $\sum_{p=1}^N \vert c_p^{(n)} \vert^2=1$. The fields $\psi_n (\ve{r})$ resulting from their propagation through the disordered system are obtained by solving the coupled dipole equations,
\begin{gather}
\Psi_n (\ve{r})=V_n(\ve{r})
-k^2  \sum_{i=1}^{N_s}  \tilde{G}_0(\ve{r},\ve{r}_i,\omega)\tilde{\alpha} (\ve{r}_i, \omega)\tilde{\psi}_i,
\label{EqPropagation1}
\\
\tilde{\psi}_i = V_n(\ve{r})
 -k^2 \sum_{\substack{j=1\\j\neq i}}^{N_s} \tilde{G}_0(\ve{r}_i,\ve{r}_j,\omega) \tilde{\alpha} (\ve{r}_j, \omega) \tilde{\psi}_j.
\label{EqPropagation2}
\end{gather}
Figure~\ref{fig:maps_T} displays the intensity patterns $I_n(x, y) = \vert \Psi_n(x, y) \vert^2$ and the corresponding integrated intensity profiles $I_n(x) = \int_0^W \mathrm{d}y \, I_n(x, y)$, shown for both the diffusive (top row) and localized (bottom row) regimes. Typical input states $V_n$ are illustrated for low transmission in the left column and high transmission in the right column. To improve clarity, these simulations are conducted for square systems ($k_0 W = k_0 L = 60$). In each regime, however, the number of scatterers is adjusted so that the distribution of transmission eigenvalues qualitatively resembles that in Fig.~\ref{fig:eigen_transm} in the diffusive (blue histogram) and localized (orange histogram) regimes.

In the diffusive regime, the intensity maps $I_n(x,y)$ are statistically invariant along the $y$-direction (see Fig.~\ref{fig:maps_T}, top), with a profile $\left< I_n(x)\right>$ that depends on the eigenvalue $T_n$. The most open channels ($T_n\simeq 1$) have a symmetric bell-shape [blue dashed line in Fig.~\ref{fig:maps_T}b)], whose maximum increases linearly with the optical thickness $L/\ell(\omega)$. For $L\gg \ell(\omega)$, their profile is well captured by the formula
\be
\left< I_{T=1}(x)\right> \simeq \left< I_{T=1}(0)\right>\left[ 1+ \frac{\pi}{2} \frac{L}{\ell(\omega)} \frac{x(L-x)}{L^2}\right],
\label{EqOpen}
\ee
where $\left< I_{T=1}(0)\right>=\left<\sum_{p=1}^N \vert c_p^{(n=1)} \vert^2 k/k_p \right> \gtrsim1$,
in agreement with the results of simulations performed in non-resonant disordered media~\cite{davy_15}. On the other hand, by extrapolating the results of Ref.~\cite{davy_15} at small transmission, we find that closed channels ($T_n \ll 1$) are expected to decay exponentially as
\be
\left< I_{T_n \ll 1}(x) \right> \simeq 2\left< I_{T=1}(x)\right> e^{-2x/\xi_n}, 
\label{EqClosed}
\ee
with $\xi_n\simeq 2L/\mathrm{ln}(4/T_n)$. This prediction is in qualitative agreement with our results for resonant media [black dash-dotted line in Fig.~\ref{fig:maps_T}a)]. Both results~\eqref{EqOpen} and~\eqref{EqClosed} drastically differ from the linear decay of plane wave propagation, $\left< I(x)\right> \varpropto (L-x)/L$, imposed by conservation of the mean current. 

In the localized regime, the intensity maps $I_n(x, y)$ exhibit significant fluctuations. Notably, the states associated with the largest transmission eigenvalues lose their invariance along the $y$-direction [see Fig.~\ref{fig:maps_T}d)]. Each of these states arises from the efficient excitation of a cluster of a few exponentially localized quasi-normal eigenmodes of the system~\cite{pena_2014, davy_2018}, enabling light to percolate through to the transmission side, with transmission potentially orders of magnitude larger than the transmission of a simple plane wave. The intensity of these sates inside the medium can also be much larger than the intensity of the open channels found in the diffusive regime (compare the intensity values in the two maps on the right). This could be attributed to a difference in the value of $\ell(\omega)$ in both scenarios. The vast majority of the transmission eigenstates present however a very low transmission. The corresponding intensity profiles $I_n(y)$ decay exponentially within the medium [see Fig.~\ref{fig:maps_T}c)], with localization lengths $\xi_n \simeq n (\pi/2) \ell(\omega)$ related to the transmission by $T_n = 4e^{-2L/\xi_n}$~\cite{beenakker_97}, similar to the behavior observed for closed channels in the diffusive regime [see Eq.~\eqref{EqClosed}].

\section{Dwell-time operator}
\label{dwell_time_matrix}

\subsection{Electromagnetic energy and the dwell-time operator}

To establish the expression of a dwell-time operator in a resonant medium, it is useful to formally write the coupled dipole equations~\eqref{EqCoupleDipole3} and~\eqref{EqCoupleDipole4} as a wave equation, $[\nabla^2 +k^2\varepsilon(\ve{r}, \omega)+i0^+]G(\ve{r}, \ve{r'}, \omega)=\delta(\ve{r}- \ve{r'})$. By noting that the exciting fields $\tilde{\psi}_i$ in Eq.~\eqref{EqCoupleDipole4} can be written as  $\tilde{\psi}_i=G(\ve{r}_i, \ve{r'}, \omega)/[1-k^2\tilde{G}_0(\ve{r}_i,\ve{r}_i,\omega)\tilde{\alpha}(\ve{r}_i, \omega)]$, we find that $G(\ve{r}, \ve{r'}, \omega)$ in Eq.~\eqref{EqCoupleDipole3} obeys a wave equation  with a frequency-dependent dielectric constant of the form
\be
\varepsilon(\ve{r},\omega) =1+ \sum_{i=1}^{N_s} \frac{\tilde{\alpha}(\ve{r}_i, \omega)}{1-k^2\tilde{G}_0(\ve{r},\ve{r}_i,\omega) \tilde{\alpha}(\ve{r}_i, \omega)} \delta(\ve{r}-\ve{r}_i),
\ee
which satisfies $\mathrm{Im}[\varepsilon(\ve{r},\omega)] = 0$, although $\mathrm{Re}[\tilde{G}_0(\ve{r}_i,\ve{r}_i,\omega)]$ diverges, since $\mathrm{Im}[1/\tilde{\alpha} (\ve{r}_i, \omega)]=k^2\mathrm{Im}[\tilde{G}_0(\ve{r}_i,\ve{r}_i,\omega)]$ in the absence of absorption. For such non-absorbing medium, the local electromagnetic energy density, associated to an arbitrary field $\psi(\ve{r}, \omega)$ solution of the wave equation, is~\cite{landau} 
\be
u(\ve{r},\omega) = \frac{\varepsilon_0}{4\omega}\partial_{\omega} [\omega^2 \varepsilon(\ve{r},\omega)] |\psi(\ve{r},\omega)|^2.
\label{EqEnergyDensity}
\ee
In Ref.~\cite{durand_19}, it was demonstrated that the total energy within the scattering region $\mathcal{V}$, given by $U(\omega)=\int_\mathcal{V} \mathrm{d}\ve{r} \,u(\ve{r}, \omega)$ where $u(\ve{r}, \omega)$ takes the form of Eq.~\eqref{EqEnergyDensity}, can be expressed in terms of the so-called dwell-time operator $ \mathbf{Q}_d(\omega)$,
\be
U (\omega)= \phi_{\mathrm{in}} \langle \psi_{\mathrm{in}} | \mathbf{Q}_d(\omega)|\psi_{\mathrm{in}} \rangle.
\label{EqUasQd}
\ee
Here, $\phi_{\mathrm{in}}$ is the power carried by the incoming field $|\psi_{\mathrm{in}} \rangle$, normalized such that $ \langle \psi_{\mathrm{in}} | \psi_{\mathrm{in}} \rangle =1$ (the original incident field must be multiplied by $\sqrt{\varepsilon_0 c/2\phi_{\mathrm{in}}}$ to get $ \psi_{\mathrm{in}}$). Equation~\eqref{EqUasQd} shows that $\tau (\omega)=U (\omega)/ \phi_{\mathrm{in}}=\langle \psi_{\mathrm{in}} | \mathbf{Q}_d(\omega)|\psi_{\mathrm{in}} \rangle$ represents the dwell-time of the field inside the scattering region. 

Focusing on the case where $|\psi_{\mathrm{in}} \rangle$ propagates from left to right (see Fig.~\ref{fig:system}) and contains no evanescent components, $|\psi_{\mathrm{in}} \rangle$ can be represented in the basis of the $N$ propagating channels. The $N \times N$ dwell-time operator takes the form~\cite{durand_19}
\be
\mathbf{Q}_d(\omega)=\mathbf{Q}(\omega)+\mathbf{Q}_i(\omega)+\mathbf{Q}_e(\omega),
\label{EqDefQd}
\ee
where $\mathbf{Q}(\omega)$ is the Wigner-Smith operator, 
\be
\mathbf{Q}(\omega)= -i\left[\mathbf{t}(\omega)^{\dagger}\partial_{\omega}\mathbf{t}(\omega)+\mathbf{r}(\omega)^{\dagger}\partial_{\omega}\mathbf{r}(\omega)
\right].
\ee
For standard input wavefronts $|\psi_{\mathrm{in}} \rangle$, such as plane waves close to normal incidence, $\mathbf{Q}(\omega)$ is responsible for the largest contribution to the dwell-time. The second term $\mathbf{Q}_i(\omega)$ results from the interference of the incident  and reflected fields. It reads
\be
\mathbf{Q}_i(\omega)=-i
\frac{
\ve{D}(\omega) \mathbf{r}(\omega) - \mathbf{r}(\omega)^\dagger\ve{D}(\omega)}
{2},
\ee 
where $\ve{D}(\omega)$ is a $N\times N$ diagonal matrix with elements $D_{mm}(\omega)= \omega/[\omega^2-(m\pi c/W)^2]$ ($m\le N$). The contribution of $\mathbf{Q}_i(\omega)$ is significant for states of low dwell-time, which do not penetrate deeper than a few mean free paths inside the scattering volume.
Finally $\mathbf{Q}_e(\omega)$ captures dwell-time contributions due to scattering into evanescent channels of the empty waveguide,
\be
\mathbf{Q}_e(\omega) = \frac{\mathbf{t}_e(\omega)^\dagger\ve{D}_e(\omega) \mathbf{t}_e(\omega)+\mathbf{r}_e(\omega)^\dagger\ve{D}_e(\omega) \mathbf{r}_e(\omega)}{2}.
\ee 
Here, $\mathbf{t}_e(\omega)$ and $\mathbf{r}_e(\omega)$ are the transmission and reflection matrices that connect the $N$ input propagation channels to $N_e$ output evanescent channels; they are of size $N_e \times N$ and take the same form as in Eqs.~\eqref{EqTM} and ~\eqref{EqRM}, except that $k_m=ik\sqrt{(m\pi /kW)^2-1}$, with $N<m\le N_e+N$. In addition, $\ve{D}_e(\omega)$ is a  $N_e\times N_e$ diagonal matrix with elements $D_{e\, mm}(\omega)= \omega/[\omega^2-(m\pi c/W)^2]$ ($N<m\le N_e+N$). The contribution of $\mathbf{Q}_e(\omega)$ to the dwell-time is negligible, except when the transverse system size causes a geometrical resonance due to the birth of a new propagating mode, \textit{i.e} when $kW$ is close to a $\pi$-integer. 

\subsection{Dwell-time eigenvalue distribution}

\begin{figure*}[t]
\centering
\includegraphics[width = 0.95\linewidth]{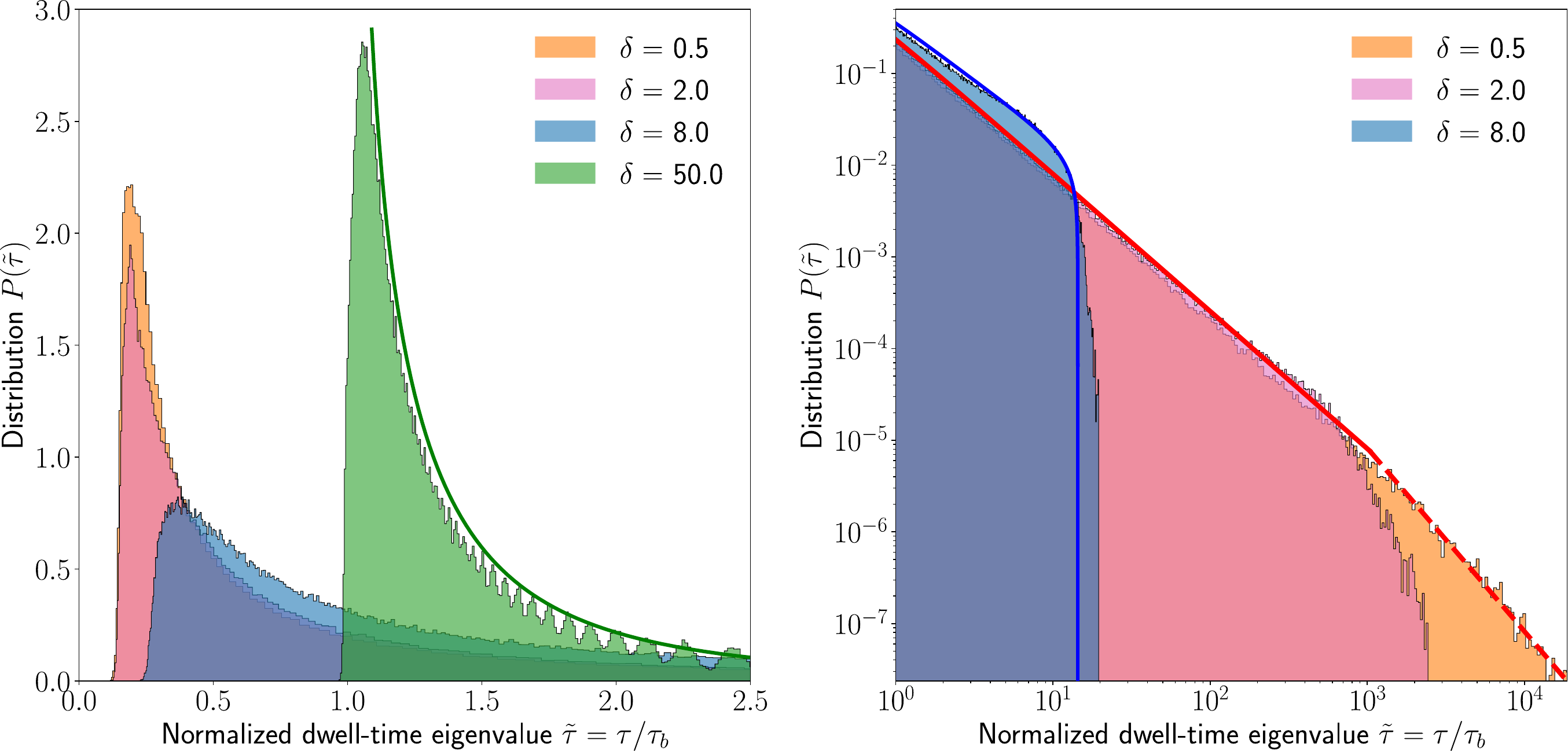}
\caption{
Probability distribution of the eigenvalues $\tau_n$ of the dwell-time operator $\mathbf{Q}_d$. Different histograms correspond to varying detuning values $\delta$, with system dimensions and density matching those used in Fig.~\eqref{fig:eigen_transm} (see Table \ref{tab:ex} for details on the propagation regimes). The left panel displays the short-time histograms on a linear scale, while the right panel shows the long-time tails on a log-log scale. The green solid line is a theoretical prediction based on the ballistic time of rays in the sample, as expressed in Eq.~\eqref{EqDTball}. The blue solid line represents the theoretical prediction for the diffusive regime, as described by the combination of Eqs.~\eqref{EqDTdiff},~\eqref{eq:tau_s_and_tau_average},~\eqref{EqMeanFreePath}, and~\eqref{EqVEnergy}. The red solid and dotted lines correspond to Eqs.~\eqref{EqDTdiff2} and~\eqref{EqDTLoc}, respectively. All times are normalized by the ballistic time, $\tau_b = L/c$.
}
\label{fig:eigen_DT}
\end{figure*}

The dwell-time eigenvalue distribution is defined as
\be
P(\tau)=\frac{1}{N} \left< \sum_{n=1}^N \delta(\tau-\tau_n)\right>,
\ee
where $\tau_n$ are the eigenvalues of the matrix $\ve{Q_d}(\omega)$ introduced in Eq.~\eqref{EqDefQd}. This distribution is shown in Fig.~\ref{fig:eigen_DT} for different detuning values $\delta$, while keeping all other parameters (the waveguide length $L$, width $W$, number of resonators $N_s$, and quality factor $Q$) fixed. To facilitate comparison between the results for $P(\tau)$ and $P(T)$, these parameters are identical to those used in Fig.~\ref{fig:eigen_transm}.

In the quasi-ballistic regime ($\delta =50$, green histogram in Fig.~\ref{fig:eigen_transm}), the distribution $P(\tau)$ reflects the dispersion of the channels of the empty waveguide. The dwell-time of channel $n$ of the waveguide is simply $\tau_n=\tau_b/\sqrt{1-(n\pi/kW)^2}$, with $\tau_b=L/c$  the ballistic time for straight propagation along the waveguide axis. In the limit of a large number of propagation channels, $N = \lfloor kW/\pi \rfloor \gg 1$,  we deduce the following distribution of normalized dwell-time $\tilde{\tau}=\tau/\tau_b$,
\be
P(\tilde{\tau})=\frac{1}{\tilde{\tau}^2\sqrt{\tilde{\tau}^2 -1 }}.
\label{EqDTball}
\ee
This simple prediction appears in good agreement with simulations, provided a scaling factor is introduced to account for the residual scattering at short dwell-time (green solid line in Fig.~\ref{fig:eigen_DT}). In the tail of the distribution, the spacing between successive $\tau_n$ becomes larger than the disorder-induced fluctuations around each $\tau_n$ value, revealing the underlying discrete nature of the channels in $P(\tilde{\tau})$. This effect disappears as $W$ increases (result not shown).

In the diffusive regime ($\delta = 8$), the distribution $P(\tilde{\tau})$ broadens, with a minimum at $\tau \ll \tau_b$ (Fig.~\ref{fig:eigen_DT}, left) and a tail extending to large times $\tau \gg \tau_b$ (Fig.~\ref{fig:eigen_DT}, right). The eigenstates associated with short dwell times correspond to reflected wavefronts, while those with long dwell times correspond to transmitted wavefronts. In Ref.~\cite{durand_19}, it is proved that the distribution $P(\tilde{\tau})$ of non-resonant diffusive media takes the form
 \be
P(\tilde{\tau}) = \frac{2\tau_s}{\pi \tau_b}\frac{1}{\tilde{\tau}^2} \sqrt{\left(\alpha \tilde{\tau}-1\right )\left(1-\beta\tilde{\tau}\right)}\left(1+\gamma \tilde{\tau}\right).
\label{EqDTdiff}
\ee
Here, $\tau_s$ is the scattering time (which expression will be given later), and $\alpha$, $\beta$ and  $\gamma$ are coefficients  that depend on  $\tau_s/\tau_b$ and $\tau_s/\langle \tau \rangle$, where $\langle \tau \rangle$ is the mean dwell time. Defining $X = \sqrt{\beta/\alpha}$,  these coefficient are solutions of  $(1-X)^2(3+2X+2X^2)/(2X(1+X^2))=\langle \tau \rangle/\tau_s$, $\alpha =(\tau_b/\tau_s) (1+X)/(1-X^2)^2$ and $\gamma = 2(\tau_b/\tau_s)X^2/(1-X^2)^2$. 
In the regime of large optical thickness, the distribution~\eqref{EqDTdiff} predicts that nonzero eigenvalues lie in the interval $\tau_s \lesssim  \tau \lesssim 4 \left< \tau\right>^2/\tau_s$, and that the maximum is reached for $\tau \simeq 4\tau_s/3$. 

In non-resonant disordered media, the scattering time and mean time are given by  $\tau_s=(\pi/2)\ell/ v_\varphi$ and  $\left< \tau \right>=(\pi/2)L/ v_\varphi$, where $v_\varphi = c/n$ is the phase velocity, $n$ being the real part of the effective refractive index experienced by the mean field~\cite{durand_19}. This implies that the largest accessible dwell-time is $\tau_{\mathrm{max}} = (2\pi/9)L^2/\ell v_\varphi= (\pi^3/9)\tau_{\mathrm{Th}}$, where $\tau_{\mathrm{Th}}=L^2/\pi^2 D_B$ is the Thouless time. Here $D_B=\ell v_E/2$ is the diffusion coefficient of the mean intensity, with $v_E=v_\varphi$ for non-resonant media, and $\tau_{\mathrm{Th}}$ is the longest mode lifetime of the corresponding diffusion equation. 

It is worth noting that the largest dwell-time $\tau_{\mathrm{max}} \sim \tau_{\mathrm{Th}}$ is significantly greater than the dwell-time $\tau_{\mathrm{PW}}$ associated with a plane wave or a random wavefront, as the latter scales with the average dwell-time $\left< \tau \right>$~\cite{durand_19}. This difference arises because $\tau_{\mathrm{PW}}$ is essentially the product of the transmission diffusive time $\tau_{\mathrm{Th}}$ and the probability $\sim \ell / L$ of reaching the transmission side. Wavefront shaping allows an increase in the transmission probability, resulting in $\tau_{\mathrm{max}} / \tau_{\mathrm{PW}} \sim L / \ell \gg 1$. Notably, this property is also preserved in resonant materials, as we shall see.

\begin{figure}
\centering
\includegraphics[width = 0.9\linewidth]{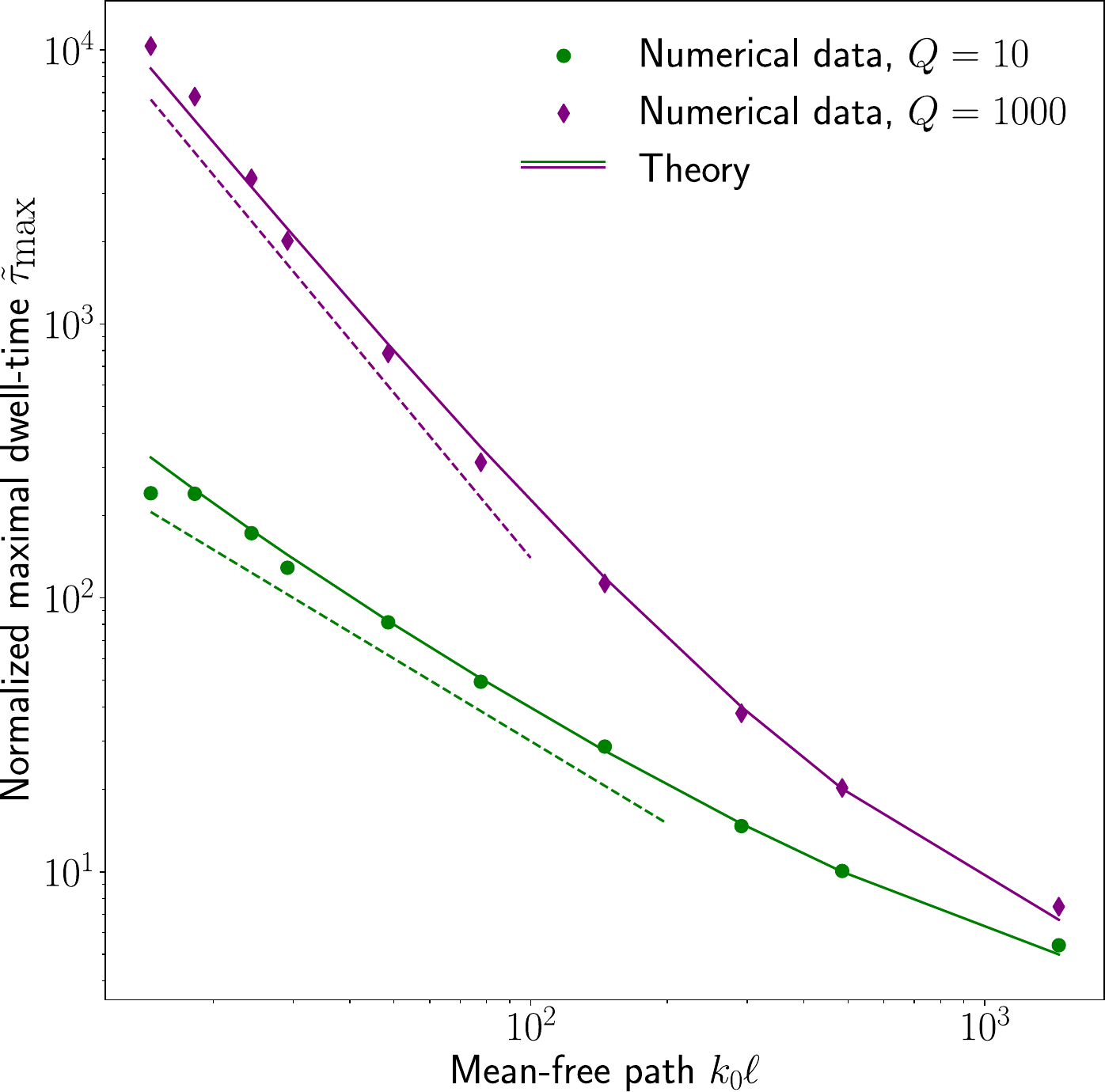}
\caption{
Maximal dwell-time in the diffusive regime for moderate (green) and large (purple) values of the quality factor $Q$ of the resonators. The normalized mean free path $k_0\ell(\omega)$, given by Eq.~\eqref{EqMeanFreePath}, is tuned by varying the number of scatterers from $N_s = 100$ to $12750$, while keeping the frequency on resonance ($\delta=0$). The system has length $k_0L = 5000$ and supports $N=47$ transverse channels. Numerical simulation results (dots) are compared to the theoretical prediction (solid lines), determined by the upper edge $1/\beta$ of the distribution~\eqref{EqDTdiff}, with a good approximation provided by Eq.~\eqref{EqTauMaxDiff}. Dotted lines highlight the $1/\ell(\omega)$ dependence of the maximal dwell-time when $Q \ll 2k_0\ell(\omega)$ and the $1/\ell(\omega)^2$ dependence when $Q \gg 2k_0\ell(\omega)$.
}
\label{fig:tau_max}
\end{figure}

In resonant disordered media, the energy velocity $v_E$ is strongly reduced near resonance, as light can spend a time $\sim \Gamma_0^{-1}$ inside each resonator between two scattering events separated by a distance $\ell(\omega)$. By adapting expressions found in Ref.~\cite{lagendijk_96} to the case of 2D scattering, we find that, in the dilute regime $k_0a \gg 1$, with $a = (N_s/WL)^{-1/2}$ the mean distance between scatterers, the energy velocity is given by
\be
\frac{1}{v_E(\omega)} = \frac{v_\varphi}{c^2}+ 
\frac{1}{\Gamma_0 \ell(\omega)} 
\left( 1+\frac{2}{\pi Q} \right),
\label{EqVEnergy}
\ee
for $\delta \ll Q$, where $v_\varphi \simeq c \left(1 + 2\delta/[(k_0a)^2(\delta^2 + 1)]\right)$.  Consequently, both the Thouless time $\tau_{\mathrm{Th}}$ and the mean dwell time $\left< \tau \right>$~\cite{pierrat_2014} are reduced near resonance. Assuming that the scaling of $\tau_{\mathrm{max}}$ with $\tau_{\mathrm{Th}}$ must hold for resonant media, we find that the scattering time and mean time are given by
\be
\tau_s = \frac{\pi}{2}\frac{\ell(\omega)}{v_E(\omega)} \ \ \mathrm{and} \ \ \langle \tau \rangle =  \frac{\pi}{2}\frac{L}{v_E(\omega)}.
\label{eq:tau_s_and_tau_average}
\ee
Very good agreement is observed in the diffusive regime between the prediction for $P(\tilde{\tau})$ given by Eqs.~\eqref{EqDTdiff},~\eqref{EqVEnergy}, and~\eqref{eq:tau_s_and_tau_average} with the above parameters and the simulation results (solid blue line in Fig.~\ref{fig:eigen_DT}, right).

An interesting result of energy velocity modulation is the potential to significantly increase the maximum accessible dwell time by enhancing the quality factor $Q$ of the resonators. To demonstrate this, we represent in Fig.~\ref{fig:tau_max} the largest normalized dwell time $\tilde{\tau}_{\mathrm{max}} = \tau_{\mathrm{max}} / \tau_b$ computed at resonance ($\delta = 0$) as a function of disorder strength, for both low and high $Q$ resonators. The disorder strength, quantified by $k\ell(\omega)$, is modified by varying the number of scatterers $N_s$, while fixing the system size. Apart from the excellent agreement between simulations and theory, we find that the dwell time in high-$Q$ systems can be enhanced by a factor of $\sim Q / 2k\ell(\omega)$ compared to low-$Q$ systems. Indeed, by considering the limit $L \gg \ell(\omega)$ in Eq.~\eqref{EqDTdiff} and neglecting the small impact of the refractive index on phase velocity, we find that the largest normalized dwell time $\tilde{\tau}_{\mathrm{max}} = \tau_{\mathrm{max}} / \tau_b$ is given by 
\be
\tilde{\tau}_{\mathrm{max}} \simeq \frac{2\pi}{9} \frac{L}{\ell(\omega)} \left[1 + \frac{Q}{2k\ell(\omega)} \right].
\label{EqTauMaxDiff}
\ee
Whereas $\tilde{\tau}_{\mathrm{max}}$ scales as $1/\ell(\omega)$ for low-$Q$ systems, it scales as $Q / \ell(\omega)^2$ for high-$Q$ systems (see dashed line in Fig.~\ref{fig:tau_max}).

Figure~\ref{fig:eigen_DT} also presents results for $P(\tilde{\tau})$ in the localized regime $\xi \lesssim L$ ($\delta = 0.5$, orange histograms). In this regime, we must distinguish between states that explore a region smaller than the localization length $\xi$ before escaping and those that stay longer and are affected by the localization process. For $\tau \lesssim \xi^2 / D_B \sim N^2 \tau_s$, states explore the system diffusively. Taking the limit $L \to \infty$ in Eq.~\eqref{EqDTdiff}, we obtain $\alpha = 1$, $\beta = 0$, and $\gamma = 0$, yielding
\be
P(\tau) = \frac{2\tau_s}{\pi} \frac{1}{\tau^2} \sqrt{\frac{\tau}{\tau_s} - 1} \quad \mathrm{for} \quad \tau \lesssim N^2 \tau_s.
\label{EqDTdiff2}
\ee
This prediction is shown as a solid red line in Fig.~\ref{fig:eigen_DT}, right. For $\tau \gtrsim N^2 \tau_s$, the distribution $P(\tau)$ can be derived by noting that the longest dwell time corresponds to the Wigner time in the localized regime, $\tau_W = \sum_{n=1}^N \tau_n \simeq \mathrm{max}(\tau_n)$. By extrapolating known results for the Wigner time in non-resonant media~\cite{texier_99, ossipov_00, bolton_99} to resonant cases, we find
\be
P(\tau) = \frac{(\pi / 2) N \ell(\omega)}{v_E(\omega) \tau^2} \quad \mathrm{for} \quad \tau \gtrsim N^2 \tau_s,
\label{EqDTLoc}
\ee
which becomes frequency-independent for high-$Q$ systems, $P(\tau) \simeq (\pi / 2) N / \Gamma_0 \tau^2$. Up to a prefactor, Eq.~\eqref{EqDTLoc} aligns with the prediction of Ref.~\cite{ossipov_18}, which investigates $P(\tau)$ for non-resonant wave scattering, with waves injected from both sides of the medium. The robustness of the power-law scaling in $P(\tau)$ originates from the distribution of decay rates $\Gamma$ of the quasi-modes in localized disordered systems~\cite{texier_99, kottos_2005}. By connecting the Wigner time and the density of quasi-modes, we find that $\int_\tau \mathrm{d}\tau' P(\tau') \varpropto N \langle \tau \rangle \int^{1/\tau} \mathrm{d}\Gamma \, \Gamma P(\Gamma)$, leading to $P(\tau) \sim 1/\tau^2$ because $P(\Gamma) \sim 1/\Gamma$ in any dimension within the localized regime~\cite{kottos_2005}. The prediction~\eqref{EqDTLoc} is shown as a dashed red line in Fig.~\ref{fig:eigen_DT}, right.

\begin{figure*}
\centering
\includegraphics[width = 0.9\linewidth]{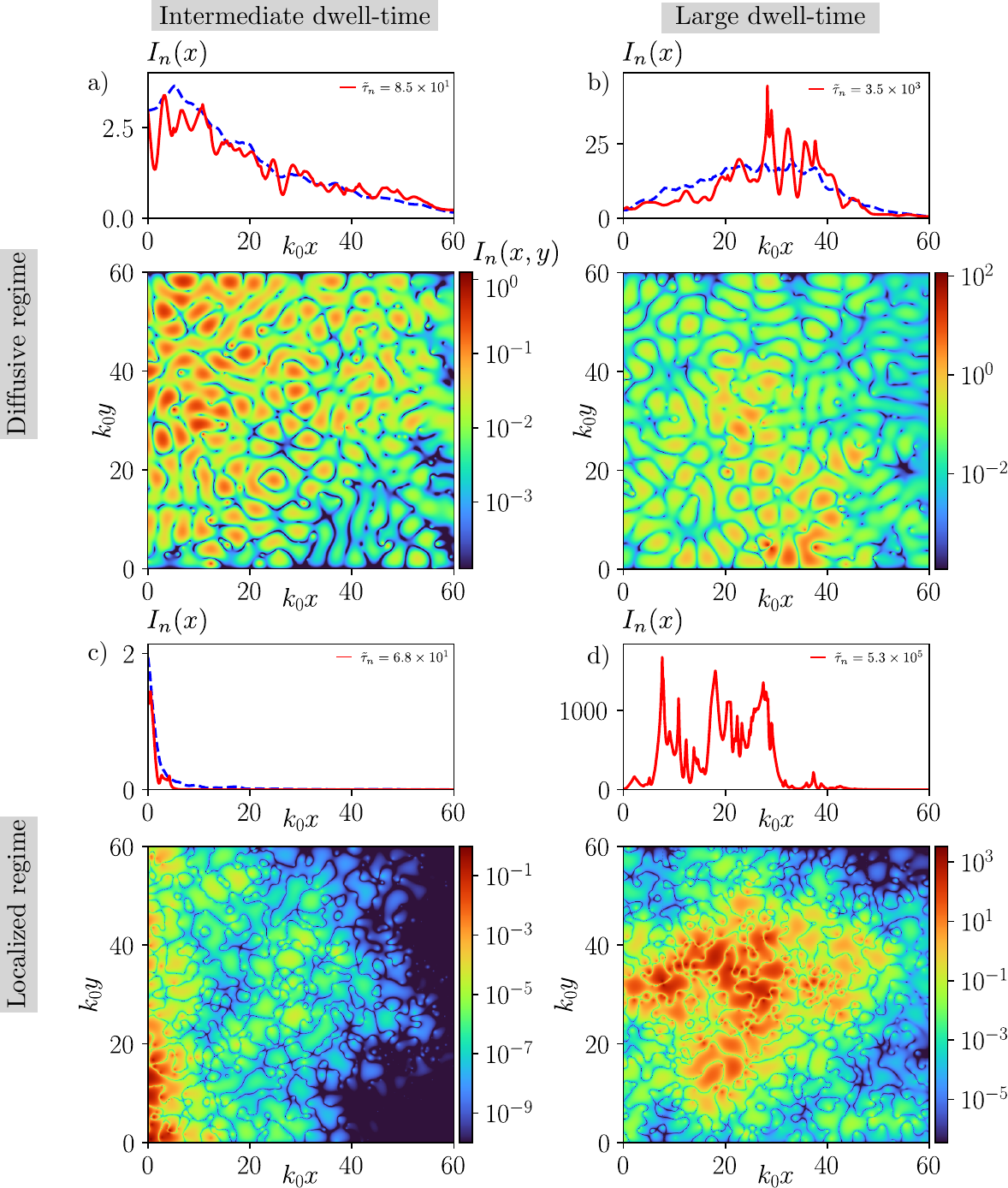}
\caption{
Intensity maps $I_n(x,y)=|\Psi_n(x, y)|^2$ and integrated intensity profiles $I_n(x)=\int_0^W \mathrm{d}y I_n(x,y)$ for typical incident wavefronts, chosen as different eigenstates $V_n$ of the matrix $\mathbf{Q}_d(\omega)$, in both the diffusive (first row) and localized regimes (second row). System dimensions ($k_0W$, $k_0L$) and resonator characteristics ($\delta$, $Q$, $N_s$) match those used in Fig.~\eqref{fig:maps_T}. The left column corresponds to the propagation of eigenstates with dwell-times $\tau_n$ close to the mean time $\left< \tau \right> = (\pi/2)\ell(\omega)/v_E(\omega) \simeq 105$ in the diffusive regime, while the right column shows eigenstates with $\tau_n$ close to the largest dwell time accessible in both regimes (see text for details). Solid red lines in $I_n(x)$ represent the specific disorder realization associated with the intensity map below, while blue dashed lines show the average intensity over 30 disorder configurations.
}
\label{fig:maps_DT}
\end{figure*}

\subsection{Eigenstate propagation}

In a symmetric manner to Fig.~\ref{fig:maps_T}, we present in Fig.~\ref{fig:maps_DT} typical intensity maps corresponding to the propagation of various eigenvectors of the dwell-time operator $\ve{Q_d}(\omega)$. These maps are obtained by solving the coupled dipole equations~\eqref{EqPropagation1} and~\eqref{EqPropagation2}, where $V_n(\ve{r})$ now represents a dwell-time eigenstate. To facilitate comparison with the intensity maps of transmission eigenstates, the system parameters ($k_0W$, $k_0L$, $N_s$, $\delta$, $Q$) are identical to those in Fig.~\ref{fig:maps_T}. This implies that the distribution of dwell-time eigenvalues (not shown) qualitatively resembles that in Fig.~\ref{fig:eigen_DT} for the diffusive (blue histogram) and localized (orange histogram) regimes, but with a larger normalized scattering time $\tau_s / \tau_b \simeq (\pi/4) Q / kL \simeq 13$, since the shorter system length ($k_0L = 60$ instead of $k_0L = 5000$) reduces the value of $\tau_b = L / c$. 
The left column of Fig.~\ref{fig:maps_DT} shows states with dwell times slightly shorter than the mean dwell time in the diffusive regime, $\langle \tau \rangle / \tau_b \simeq (\pi/4) Q / k \ell(\omega) \simeq \pi Q / (ka)^2 \simeq 105$ at $\delta = 0$, yet significantly longer than the scattering time, which corresponds to the most probable time in the diffusive regime, $4 \tau_s / 3 \tau_b \simeq 17$. Conversely, the right column of Fig.~\ref{fig:maps_DT} shows states with  dwell times close to the largest accessible values, in both the diffusive and localized regimes. In the diffusive regime, Eq.~\eqref{EqTauMaxDiff} gives $\tau_{\mathrm{max}}/\tau_b \simeq (\pi/9) Q kL/ [k \ell(\omega)]^2 \simeq (16\pi/9) Q / (ka)^2 \simeq 10^4$. In the localized regime, a rough estimate of the accessible $\tau_{\mathrm{max}}$ using $N_r$ disorder configurations is the solution of  $NN_r\int_{\tau_\mathrm{max}}^\infty \mathrm{d}\tau P(\tau)=1$, with $ P(\tau)$ given by Eq.~\eqref{EqDTLoc}. This gives  $\tau_\mathrm{max}/\tau_b \simeq  (\pi/4) Q N^2N_r/ kL \simeq 5\times 10^5$ for $N_r=100$.

In the diffusive regime (Fig.~\ref{fig:maps_DT}, upper row), dwell-time eigenstates are statistically invariant along the transverse direction, but the profile $\left< I_n(x) \right> = \int_0^W \mathrm{d}y \, \left< I_n(x,y) \right>$ depends on the eigenvalue $\tau_n$. Generally, the larger $\tau_n$, the deeper the penetration of $I_n(x)$ into the material. This property, which is apparent in non-resonant media, is less straightforward in resonant media because, according to the definition~\eqref{EqUasQd} of $\ve{Q}_d$, the normalized eigenvalue $\tilde{\tau}_n=\tau_n/\tau_b$ is related to the field $\Psi_n(\ve{r}, \omega)$ by
\be
\tilde{\tau}_n= \int_\mathcal{V} \frac{\mathrm{d}\ve{r}}{L} \, \frac{\partial_{\omega} \left[ \omega^2 \varepsilon(\ve{r}, \omega) \right]}{2\omega} |\Psi_n(\ve{r}, \omega)|^2,
\label{EqDTvsField}
\ee
which differs in a non-trivial way from the non-resonant result $ (c/v_\varphi) \int_0^L \mathrm{d}x \, I_n(x) /L $. In particular, the values of $\tilde{\tau}_n$ are much larger than the typical values of $ I_n(x)$ because most of the energy is stored inside the material degrees of freedom instead of the field.  However, the profiles found for $\left< I_n(x) \right>$ are consistent with those for non-resonant media, as reported in Ref.~\cite{durand_19}. As $\tau_n$ increases, $\left< I_n(x) \right>$ evolves from an exponential profile near the injection surface to a profile that deposits most of the intensity within the bulk of the disordered system. 
The maximum dwell-time eigenstates [Fig.~\ref{fig:maps_T}b)] exhibit a peak intensity that is slightly shifted to the left of the sample midpoint $x = L/2$, due to the asymmetry induced by left-side injection, which breaks the system's statistical mirror symmetry. This symmetry can be restored by injecting light from both sides~\cite{durand_19}. Although no theoretical prediction exists for $\left< I_n(x) \right>$, we find that the profiles are qualitatively similar to those associated with transmission eigenchannels, showing slightly more intensity inside the sample for the largest dwell-time eigenstates than for the open channels [compare Fig.~\ref{fig:maps_T}b), with Fig.~\ref{fig:maps_DT}b)].

In the localized regime (Fig.~\ref{fig:maps_DT}, lower row), eigenstates with dwell-times similar to the mean diffusive dwell-time exhibit behavior quite distinct from that of corresponding diffusive states [compare Fig.~\ref{fig:maps_DT}a) and c)]. Notably, the intensity profile $\left< I_n(x) \right>$ is significantly more localized near the injection surface, suggesting that there is not always a straightforward relationship between the dwell-time, as defined in Eq.~\eqref{EqDTvsField}, and the integrated intensity. 
In contrast, wavefronts with dwell-time $\tau \gg N^2 \tau_s$ produce intensity maps that are exponentially localized within the sample [Fig.~\ref{fig:maps_DT}d)], indicating that they can efficiently couple to quasi-modes deeply localized within the material. Interestingly, these states not only reach exceptionally large dwell-times but also generate intensity patterns that are substantially more intense than those in the diffusive regime [compare Fig.~\ref{fig:maps_DT}b) and d)], and even surpass the intensity of the largest transmission eigenchannels in the localized regime [compare Fig.~\ref{fig:maps_T}d), with Fig.~\ref{fig:maps_DT}d)]. This is because transmission eigenchannels optimize coupling between localized quasi-modes to maximize transmission, whereas dwell-time eigenchannels optimize targeting the deepest and most isolated quasi-modes.

\section{Conclusion}
\label{conclusion}

In this work, we have shown that the transmission and dwell-time eigenvalue distributions of resonant disordered media, $P(T)$ and $P(\tau)$,
can be fully characterized by two mesoscopic parameters --- the transport mean free path $\ell(\omega)$ and the energy velocity $v_E(\omega)$ --- along with two geometric parameters, the system length $L$ and the number of transverse channels $N$.

In the diffusive regime ($\ell(\omega) \ll L \ll \xi$), $P(T)$ follows the bimodal distribution~\eqref{EqBimodal}, parametrized by the optical thickness $L/\ell(\omega)$, while $P(\tau)$ follows Eq.~\eqref{EqDTdiff}, which is governed by both the optical thickness and the time between successive scattering events, $\ell(\omega)/v_E(\omega)$. In this regime, wavefront shaping is highly effective, enabling transmission and energy storage to surpass those achievable with a plane wave or random wavefront by a factor of $L/\ell(\omega) \gg 1$. Indeed, $\langle T \rangle \varpropto \ell(\omega)/L$ and $T_\mathrm{max} \simeq 1$ yield $T_\mathrm{max}/\langle T \rangle \varpropto L/\ell(\omega)$, while $\langle \tau \rangle \varpropto L/v_E(\omega)$ and $\tau_\mathrm{max} \varpropto L^2/[\ell(\omega) v_E(\omega)]$ give $\tau_\mathrm{max}/\langle \tau \rangle \varpropto L/\ell(\omega)$. Additionally, using the explicit expression for the energy velocity, we demonstrated that $\tau_\mathrm{max}$ can be enhanced by a factor $c/v_E(\omega) \simeq Q / 2k\ell(\omega) \gg 1$ when employing high-$Q$ resonators instead of low-$Q$ ones [see Eq.~\eqref{EqTauMaxDiff} and Fig.~\ref{fig:tau_max}].  This contrasts with the portion of energy stored in the field, which scales as $L/\ell(\omega)$ for both the open channels [Fig.~\ref{fig:maps_T}b)] and the eigenstates of $\ve{Q}_d(\omega)$ associated with $\tau_\mathrm{max}$ [Fig.~\ref{fig:maps_DT}b)]. Notably, the field energy of the latter appears independent of the energy velocity and therefore independent of the quality factor of the resonators, except through the detuning $\delta$.

In the localized regime ($W < \xi \simeq (\pi/2)N\ell(\omega) < L$), the distributions $P(T)$ and $P(\tau)$ depend on an additional parameter: the number of channels $N$. Most transmission eigenvalues are close to zero, while the largest transmission eigenvalue typically follows a log-normal distribution [see Eq.~\eqref{EqPMax}], exhibiting significant fluctuations [$\ln(T_\mathrm{max})$ has a standard deviation $\varpropto \sqrt{L/\xi}\gg 1$]. This implies that a plane wave or a random wavefront typically yields a transmission $\varpropto e^{-2L/\xi}/N \ll 1$, although specific disorder configurations can allow $T_\mathrm{max}$ to approach unity. The enhancement $T_\mathrm{max}/\left< T \right>$ achieved through wavefront shaping is therefore even more pronounced than in the diffusive regime. The same applies to the control of energy storage via wavefront shaping. In contrast to the diffusive regime, $P(\tau)$ remains unbounded in the localized regime [see Eq.~\eqref{EqDTLoc}]. Since the mean time $\left<\tau\right>$ is proportional to the mean density of states and remains largely unaffected by localization, we find $\tau_\mathrm{max}/\left< \tau \right> \varpropto N^2N_r \ell(\omega)/L \gg 1$, where $N_r$ is the number of disorder realizations probed. In addition, the advantage observed in the diffusive regime of working with high-$Q$ resonators is preserved in the localized regime, as $\tau_\mathrm{max}$ remains proportional to $c/v_E(\omega)$.

Previous predictions apply to a wide range of disordered structures composed of resonant units. A prototypical system could consist of sub-wavelength resonant cylinders aligned along the transverse $z$-direction. The two-dimensional configuration considered in this work can also describe more practical structures, such as thin metallic waveguides filled with small high-dielectric cylinders~\cite{Laurent_2007, Aubry_2020}. Moreover, we anticipate that our predictions in the diffusive regime will also hold for three-dimensional systems, including those made of resonant dielectric scatterers, cold atoms, or artificial atoms.

An important aspect of energy storage in resonant media not addressed in this work is the detailed analysis of how energy is partitioned between the light and material degrees of freedom. While it is evident that high-$Q$ systems store most of the energy within the material, it remains unclear to which degree this partitioning is influenced by factors such as disorder strength and the propagation regime (quasi-ballistic, diffusive, or localized). Additionally, it is unclear to which degree this partition can be controlled through wavefront shaping techniques. This issue will be explored in a forthcoming publication.

\begin{acknowledgments}

This research has been supported by the ANR project MARS\_light under reference ANR-19-CE30-0026 and by the program ``Investissements d’Avenir" launched by the French Government.

\end{acknowledgments}

\appendix

\section{Coupled dipole equations in a waveguide}\label{app:couple_dipole}

In this appendix, we derive the coupled dipole equations for an assembly of resonant scattering dipoles in a waveguide. The scalar Green's function of the wave equation is given by 
\be
G(\ve{r}, \ve{r'}, \omega)=\tilde{G}_0(\ve{r}, \ve{r'}, \omega) 
-k^2 \alpha(\omega)  \sum_{i=1}^{N_s}  \tilde{G}_0(\ve{r},\ve{r}_i,\omega)\psi_i,
\label{EqCoupleDipole1}
\ee
where 
$\alpha(\omega)$ is the free space polarizability of a resonant scatterer and $\tilde{G}_0(\ve{r}, \ve{r'}, \omega)$ is the retarded Green's function of the wave equation in the empty waveguide with Dirichlet boundary conditions in the $y-$direction, defined in Eq.~\eqref{EqGreenGuide}. In addition, the field $\psi_i$, exciting the scatterer located at the position $\ve{r}_i$, includes the contribution of the source, $\tilde{G}_0(\ve{r}_i,\ve{r}',\omega)$, as well as the contributions of the different scatterers,
\begin{align}
\psi_i = \tilde{G}_0(\ve{r}_i,\ve{r}',\omega) + \psi^\circlearrowleft(\ve{r}_i)
 -k^2 \alpha(\omega) \sum_{\substack{j=1\\j\neq i}}^{N_s}  G_0(\ve{r}_i,\ve{r}_j,\omega)\psi_j.
 \label{EqCoupleDipole2}
\end{align}
Note in particular the radiation-reaction contribution from the scatterer $i$,
\be
\psi^\circlearrowleft(\ve{r}_i)= -k^2\Delta G_0(\ve{r}_i, \omega)\alpha(\omega)\psi_i,
\ee
where $\Delta G_0(\ve{r}_i, \omega)=\tilde{G}_0(\ve{r}_i,\ve{r}_i,\omega) -G_0(\ve{r}_i,\ve{r}_i,\omega)$, and $G_0(\ve{r}, \ve{r}',\omega)$ is the retarded free space Green's function. By definition, $\psi^\circlearrowleft(\ve{r}_i)=0$ in free space (where $\tilde{G}_0=G_0$), because the effect of the radiation-reaction in that case is already encapsulated in the resonance frequency $\omega_0$ and linewidth $\Gamma_0$  of the polarizability $\alpha(\omega)$. 

By introducing the effective exciting field $\tilde{\psi}_i = \psi_i - \psi^\circlearrowleft(\ve{r}_i)$ and local polarizability $\tilde{\alpha} (\ve{r}_i, \omega)=\alpha(\omega)\psi_i/\tilde{\psi}_i$, we can rewrite the coupled dipole equations~\eqref{EqCoupleDipole1} and~\eqref{EqCoupleDipole2} as Eqs.~\eqref{EqCoupleDipole3} and \eqref{EqCoupleDipole4}, where $\tilde{\alpha} (\ve{r}_i, \omega)$ is given in Eq~\eqref{EqRenormalizedAlpha}. 
Similar renormalization procedures for the polarizability due to the environment can be found in the literature; see, for example, Refs.~\cite{novotny_12,castanie_12}.

\section{Numerical computation of $\Delta G_0(\ve{r}_i, \omega)$}
\label{app:delta_G0}

In order to explicitly calculate the renormalization contribution $\Delta G_0(\ve{r}_i, \omega)$ appearing in Eq.~\eqref{EqRenormalizedAlpha}, two strategies can be adopted. Using the Poisson summation formula, we can write it as 
\be
\Delta G_0(\ve{r}_i, \omega)=  \sum_{n\in \mathbb{Z}^*} (-1)^n G_0(\ve{r}_i, \ve{r}_i^{(n)},\omega),
\ee
where $\ve{r}_i^{(n)}$ are the positions of the mirror images of the resonator located in $\ve{r}_i $: $\ve{r}_i^{(n)} = (x_i,nW+y_i)$ if $\vert n \vert$ is even, and $\ve{r}_i^{(n)} = (x_i,(n+1)W-y_i)$ if $ \vert n \vert$ is odd. However, in practice, this sum is slowly converging. Alternatively, we may use a modal expansion of the Green's function $\tilde{G}_0$, such as the one given in Eq.~\eqref{EqGreenGuide}, but we need to properly account for the fact that both $\tilde{G}_0(\ve{r}_i,\ve{r}_i,\omega)$ and $G_0(\ve{r}_i,\ve{r}_i,\omega)$ diverge, although their difference does not. This can be done by replacing $ G_0(\Delta x, \Delta y) \equiv G_0(\ve{r}, \ve{r}')$ (with $\Delta x=x-x'$, $\Delta y=y-y'$) by its convolution with a Gaussian function of finite width $b$ along the $y$-direction, $G_0^{(b)}(\Delta x, \Delta y)=G_0(\Delta x, \Delta y) \otimes e^{-\Delta y^2/2b^2}/\sqrt{2\pi}b$, and using the relation
\begin{align}
\Delta G_0(\ve{r}_i, \omega) = &e^{\frac{k^2b^2}{2}} \sum_{n\in \mathbb{Z}} (-1)^n G_0^{(b)}(\ve{r}_i, \ve{r}_i^{(n)},\omega) 
\nonumber
\\
&-G_0^{(b)}(\ve{r}_i,\ve{r}_i,\omega),
\end{align}
which holds for $kb \ll 1$. Explicit calculation gives
\begin{align}
\Delta G_0(\ve{r}_i, \omega) = &  \sum_{n>0}\frac{e^{(k^2-\pi^2 n^2/W^2)b^2/2}}{4ik_nW}\left(1- e^{2i\pi ny_i/W} \right)
\nonumber
\\
&-\frac{\gamma + \ln(b^2k^2/8)-i\pi}{4\pi} +\mathcal{O}(kb).
\label{EqShiftMode}
\end{align}
Simulation results presented in the following sections have been obtained by computing the polarizability $\tilde{\alpha}(\ve{r}_i, \omega)$ defined in Eq.~\eqref{EqRenormalizedAlpha} with the expression~\eqref{EqShiftMode} for each scatterer position $\ve{r}_i$. Convergence is reached for a number of terms in Eq.~\eqref{EqShiftMode} of the order of $\mu N_s$, with $\mu\gg 1$.

\bibliography{biblio}
	
\end{document}